\documentclass[11pt,a4paper]{article}
\usepackage[margin=2.4cm]{geometry}
\usepackage{amsmath,amssymb,bm}
\usepackage{graphicx}
\usepackage{booktabs}
\usepackage{longtable}

\usepackage[colorlinks=true,allcolors=blue]{hyperref}
\usepackage{caption}
\newcommand{\eHOMO}{\varepsilon_{\mathrm{HOMO}}}
\newcommand{\IHOMO}{I_{\mathrm{HOMO}}}
\newcommand{\vxc}{v_{\mathrm{xc}}}
\newcommand{\Kg}{K_{\mathrm{g}}}
\newcommand{\eVh}{\mathrm{eV}^{1/2}}

\title{Vertical ionization energies from inverse generalized Kohn--Sham
calculations at fixed electron density}
\author{V.~Subramanian$^{a,b}$, B.~Kanungo$^{a}$, J.~P.~Perdew$^{c}$,
V.~Gavini$^{a,d}$ and G.~I.~Csonka$^{c,*}$}
\date{\small
$^{a}$\,Department of Mechanical Engineering, University of Michigan,
Ann Arbor, Michigan 48109, USA\\
$^{b}$\,Michigan Institute for Computational Discovery and Engineering,
University of Michigan, Ann Arbor, Michigan 48109, USA\\
$^{c}$\,Department of Physics and Engineering Physics, Tulane University,
New Orleans, LA 70118, USA\\
$^{d}$\,Department of Materials Science and Engineering, University of Michigan,
Ann Arbor, Michigan 48109, USA\\[2pt]
$^{*}$\,Corresponding author.\\[4pt]
}

\begin{document}
\maketitle

\begin{abstract}
\noindent
The exact Kohn--Sham (KS) system must have its highest occupied orbital energy
equal to the negative of the lowest vertical ionization energy (VIE). No
comparable theorem connects the deeper occupied levels $\varepsilon_k$ to the
higher VIEs $I_k$. Chong, Gritsenko and Baerends (CGB) established empirically
that $I_k + \varepsilon_k$ is proportional to
$\sqrt{\varepsilon_{\mathrm{HOMO}}-\varepsilon_k}$. We ask whether a similar
relation holds for the exact generalized Kohn--Sham (GKS) system, which
incorporates a fraction $\alpha$ of Hartree--Fock exchange. Inverting near-exact
CCSD/aug-cc-pVQZ ground-state densities of LiH, LiF, H$_2$O, HF, CO, N$_2$ and
F$_2$ onto the $\alpha$-family of one-particle operators at
$\alpha = 0, 0.25, 0.5, 0.75, 1.0$, we obtain the exact GKS systems at each
admixture. Because the density is held fixed and only the operator is varied,
the resulting occupied spectra separate the effect of exact-exchange admixture
on orbital energies from its effect on the density, two contributions
inseparable in any self-consistent hybrid calculation. The CGB relation holds
broadly across the family, with a proportionality coefficient linear in the
admixture, $c(\alpha) = c(0) - s\alpha$, where $c(0) = 1.185 \pm 0.017~\eVh$ 
and $s = 1.987 \pm 0.041~\eVh$, with $R^{2} \ge 0.9996$ for each of H$_2$O, CO,
HF and N$_2$. Using the response-potential matrix elements tabulated by 
Gritsenko, Bra\"ida and Baerends, we decompose the CGB coefficient into response
and spectroscopic parts, and find $s$ to agree with the homogeneous-electron-gas 
step-height coefficient $\Kg = 1.9932~\eVh$, which we conjecture is its origin. 
  The coefficient vanishes at $\alpha^{*} = 0.597 \pm 0.008$. At $\alpha^{*}$ the HOMO-referenced GKS occupied levels satisfy
$\varepsilon_k \approx -I_k$ across the whole occupied range. The
zero crossing is fixed by the core orbitals; in the outer valence the CGB coefficient is
already indistinguishable from zero at $\alpha=0$.
\end{abstract}

\section{Introduction}

The $i$-th vertical ionization energy (VIE) of a ground-state molecule, measured
in photoemission, is the energy required to remove an electron without relaxing
the equilibrium nuclear positions; much of the literature cited here calls the
same quantity a vertical ionization potential. We label the $N$ occupied
orbitals so that $i$ runs from $1$ for the deepest to $N$ for the highest
occupied, the ionization energy decreasing as $i$ increases. The Kohn--Sham (KS)
formulation of density functional theory~\cite{KS1965} replaces the interacting
system by non-interacting electrons in a local potential chosen to reproduce the
exact density. Its $N$-th, or highest-occupied, molecular orbital (HOMO) energy equals the
negative of the lowest VIE, the theorem of Levy, Perdew and Sahni~\cite{PPLB1982,LPS1984,AvB1985}. (The
analysis of Ref.~\cite{PPLB1982} is widely, but not universally,
accepted~\cite{Baerends2022}.) A comparable theorem relating the deeper
KS orbital energies to higher VIEs does not exist, yet accurate KS occupied
spectra reproduce the photoelectron ladder far better than Hartree--Fock (HF)
Koopmans values.

Chong, Gritsenko and Baerends (CGB) analyzed the relation between VIEs and Kohn--Sham eigenvalues in 2002~\cite{CGB2002}. Their key empirical observation, made in their paper for core orbitals, is that $I_k + \varepsilon_k$ is closely proportional to
$\sqrt{\varepsilon_N-\varepsilon_k}$. The CGB relation holds for the Kohn--Sham
system, which is governed by a local multiplicative potential. Whether such a
relation exists for the generalized Kohn--Sham (GKS) system, which admits any single-electron operator, local or nonlocal, multiplicative
or nonmultiplicative, remains an open question.
Answering it requires the solution of the inverse GKS problem of finding the
exact GKS system, that is the orbitals, eigenvalues and local multiplicative
potential, from a target density. An accurate solution to the inverse GKS
problem has until recently been lacking. This work employs the inverse GKS
approach of Subramanian \textit{et al.}~\cite{invGKS2026}, based on a
systematically convergent finite-element basis, to investigate whether a
CGB-like relation holds for the GKS system.

We use the most common form of the GKS system, which adds a fraction $\alpha$ of
the nonlocal Hartree--Fock exchange operator to a local multiplicative
potential. Holding the density fixed at the coupled-cluster singles and doubles
(CCSD) result and inverting onto the $\alpha$-family of GKS operators separates
two effects that are entangled in every self-consistent hybrid calculation: what
a functional does to the density, and what its one-particle operator does to the
spectrum of a given density. The first is switched off by construction, so that
the inversion isolates the effect of the GKS operator on the occupied spectrum.
We find a CGB-like relation for the GKS systems, with an $\alpha$-dependent
proportionality coefficient, and we find that coefficient to approach zero near $\alpha^{*} \approx 0.6$ for each of the five molecules fitted. In other words, at $\alpha = \alpha^{*}$ the HOMO-referenced GKS occupied levels
satisfy $\varepsilon_k \approx -I_k$ across the whole occupied range.

The admixture parameter itself originates with Becke, who introduced the
mixing of exact and semilocal exchange and fixed the fraction at one half on
adiabatic-connection grounds~\cite{Becke1993a}, then reduced it to $0.20$ by
fitting thermochemical data~\cite{Becke1993b}; the case for hybrid methods as
it stood shortly afterwards is set out in Ref.~\cite{KBP1996}. The generalized
Kohn--Sham framework, in which such a nonlocal operator is an exact realization
of density-functional theory, was formulated later~\cite{Seidl1996,Garrick2020}.
Recently, non-self-consistent global meta-GGA hybrids have been
proposed~\cite{Gopidi2026}, with a fraction $0.1$ of exact exchange in the total
energy and $0.5$ in the effective potential, to minimize both the
functional-driven and density-driven errors of the total energy. The value
$\alpha^{*}$ reported below is an independent determination of the same
parameter, obtained from photoelectron spectra at fixed exact density rather
than from thermochemistry. This paper is dedicated to the memory of Axel Becke
(1953--2025). Some reminiscences of discussions between Becke and one of the
present authors (JPP) appear in another Becke special issue~\cite{Perdew2026}.

The paper is organized as follows. Section~\ref{sec:theory} defines the GKS
family at fixed density, states the CGB relation and the two-parameter form we
fit, and introduces the HOMO-referenced deviation used throughout.
Section~\ref{sec:comp} gives computational details and the provenance of every
experimental VIE. Section~\ref{sec:res} validates the inversion against
literature exact-KS eigenvalues, examines the $\alpha=0$ error floor, reports the fitted CGB coefficient and its linearity in
$\alpha$, and shows that the core orbitals fix its zero crossing. Sections~\ref{sec:disc}--\ref{sec:concl} give the
discussion, the copper oxide outlook and the conclusions.

\section{Theory}\label{sec:theory}

\subsection{The GKS family at fixed density}\label{sec:gksfamily}

For a fixed target density $\rho_0$ and a chosen exact-exchange fraction
$\alpha$ we seek the local multiplicative remainder potential
$\delta \vxc^{\alpha}$ such that the ground state of
\begin{equation}
\hat{h}^{\alpha}
= -\tfrac{1}{2}\nabla^{2} + v_{\mathrm{ext}} + v_{\mathrm{H}}[\rho_0]
+ \alpha\,\hat{\Sigma}_{\mathrm{x}} + \delta \vxc^{\alpha}
\label{eq:gks}
\end{equation}
reproduces $\rho_0$ exactly. Here $\hat{\Sigma}_{\mathrm{x}}$ is the nonlocal
Fock exchange operator built from the occupied orbital eigenstates of
$\hat{h}^{\alpha}$ itself, so the inversion is nonlinear. At $\alpha=0$ this
reduces to the standard KS inversion and $\delta \vxc^{0} = \vxc$. At
$\alpha=1$ exchange has been transferred entirely into the operator, so the
multiplicative remainder carries no exchange contribution and is a
correlation-only potential. It is not the Kohn--Sham correlation potential
$v_{\mathrm{c}} = \vxc - \delta E_{\mathrm{xx}}/\delta\rho_0(\mathbf{r})$,
where $E_{\mathrm{xx}}$ is the exchange energy defined using the Kohn--Sham
orbitals. The orbitals that diagonalize $\hat{h}^{\alpha}$ reproduce $\rho_0$
and are neither Kohn--Sham nor Hartree--Fock orbitals, so the remainder is
correlation defined relative to the Fock reference built from those
orbitals. It should not be confused with a multiplicative
exchange potential of the optimized-effective-potential type, which is an
$\alpha=0$ object: once exchange is represented by $\hat{\Sigma}_{\mathrm{x}}$
it cannot also appear in the remainder without being counted twice.
Consistently, the remainder at $\alpha=1$ carries no $-1/r$ tail.

Every member of the family is exact for its own auxiliary system: each
reproduces $\rho_0$. The eigenvalues $\varepsilon_i^{\alpha}$, however,
differ, because they are eigenvalues of different operators.

\subsection{The CGB relation and its extension}\label{sec:cgb}

CGB's~\cite{CGB2002} central formal result is a set of exact linear relations between the
$\varepsilon_i$ and the ionization energies $I_j$,
\begin{equation}
\mathbf{M}\bm{\varepsilon} = -\mathbf{P}\mathbf{I} + \bm{\varepsilon}^{\mathrm{resp}},
\qquad
\bm{\varepsilon} = -\mathbf{M}^{-1}\mathbf{P}\mathbf{I}
+ \mathbf{M}^{-1}\bm{\varepsilon}^{\mathrm{resp}},
\label{eq:cgbexact}
\end{equation}
with
\begin{equation}
M_{ij} = 2\!\int\! \frac{|\phi_i(\mathbf{r})|^{2}|\phi_j(\mathbf{r})|^{2}}{\rho(\mathbf{r})}\,d\mathbf{r},
\qquad
P_{ij} = 2\!\int\! \frac{|\phi_i(\mathbf{r})|^{2}|d_j(\mathbf{r})|^{2}}{\rho(\mathbf{r})}\,d\mathbf{r}.
\end{equation}
Here the $\phi_i$, $\varepsilon_i$ and $\rho$ are the occupied Kohn--Sham
orbitals, their eigenvalues and the electron density of the neutral molecule,
while the $d_j$ are the Dyson orbitals and
$\bm{\varepsilon}^{\mathrm{resp}}$ collects the expectation values, for a
Kohn--Sham orbital, of a response potential (the step-structure part of the
exchange--correlation potential) defined by CGB. The two sets of numbers are of
different kinds. The $I_j$ are properties of the interacting molecule, measured
in photoemission and independent of any calculation, whereas the
$\varepsilon_i$ are eigenvalues of an auxiliary non-interacting system. Only
$\varepsilon_N$ is tied to an ionization energy by a theorem; the relation of
the rest to the photoelectron ladder is the subject of this paper.

More relevant to our work is CGB's separate empirical observation that the
deviation of $\varepsilon_k$ from $-I_k$ is closely proportional to
$\sqrt{\varepsilon_N-\varepsilon_k}$ for core orbitals. We connect the proportionality coefficient to the step heights $w_i$ of the
response potential. Reference~\cite{GLLB1995} writes the coefficient as a density
functional $K[\rho]$ (its Eq.~23) and evaluates it for the homogeneous electron
gas (its Eq.~32), which gives
\begin{equation}
w_i = \Kg\,\sqrt{\mu-\varepsilon_i}, \qquad \mu = \varepsilon_N, \qquad
\Kg = \frac{8\sqrt{2}}{3\pi^{2}} = 0.38211~\mathrm{hartree}^{1/2} = 1.9932~\eVh ,
\label{eq:stepheight}
\end{equation}
the origin of the value $1.99~\eVh$ quoted in Ref.~\cite{CGB2002}. A later analysis by Gritsenko, Bra\"ida and
Baerends~\cite{GBB2003} decomposed the relation further using Dyson orbitals,
and it is that decomposition which we use in Sec.~\ref{sec:disc}.

As an extension to GKS systems, we write the deviation of $-I_k$ from $\varepsilon_k^{\alpha}$ as
\begin{equation}
I_k + \varepsilon_k^{\alpha}
   = c(\alpha)\,\sqrt{\eHOMO^{\alpha}-\varepsilon_k^{\alpha}} + b(\alpha)
\label{eq:fit}
\end{equation}
and report $c(\alpha)$ both for the full occupied spectrum and for the
outer-valence orbitals alone, as the two fits give different answers. We use 
$\varepsilon_N$ and $\eHOMO$ interchangeably. The
ionization energies carry no $\alpha$, being properties of the molecule, whereas
$\varepsilon_k^{\alpha}$ is a property of the GKS auxiliary system; the fit is
performed separately at each $\alpha$, and $c(\alpha)$ and $b(\alpha)$ are its
two parameters.

In an inverse GKS or inverse KS calculation $\eHOMO$ is extremely sensitive to
the far-field decay of the target density. The target densities here are
obtained in a Gaussian basis, which decays as a Gaussian rather than
exponentially, so it is numerically difficult to attain good agreement between
$\eHOMO$ and $-\IHOMO$, which the theorem of Levy, Perdew and
Sahni~\cite{PPLB1982,LPS1984,AvB1985} requires to be equal. Throughout, $\eHOMO$ is the eigenvalue of the orbital carrying the first
ionization, fixed across $\alpha$. In N$_2$ at $\alpha=1$, that orbital ($3\sigma_g$) lies 0.08 eV below the $1\pi_u$ pair (which is now the level that makes the abscissa in Eq. (5) zero). To remove any such numerical error in $\eHOMO$ from the
analysis, we shift all eigenvalues to align $\eHOMO$ with $-\IHOMO$. That is, we replace the left-hand side of
Eq.~\eqref{eq:fit} by
\begin{equation}
D_k(\alpha) = I_k + \varepsilon_k^{\alpha} - (\IHOMO + \eHOMO^{\alpha}),
\label{eq:Dk}
\end{equation}
which we call the HOMO-referenced deviation. The shift is a constant within a
molecule and the abscissa (the square root in Eq.~\eqref{eq:fit}) depends only on
differences of eigenvalues, so the
shift affects the intercept $b(\alpha)$ and not the slope $c(\alpha)$. The same shift was used in Ref.~\cite{GBB2003}, where GGA-BP orbital energies shifted by
$\eHOMO + \IHOMO$ recovered the valence VIEs to $0.09$~eV average deviation,
against $0.08$~eV for the exact KS spectrum.

\section{Computational details}\label{sec:comp}

Ground-state densities were computed at the CCSD/aug-cc-pVQZ level for LiH, LiF,
H$_2$O, CO, N$_2$, HF and F$_2$. The geometries are those of Ref.~\cite{Mezei2017}, and their bond lengths and angles
are given in Table~S1. These seven molecules are drawn from the closed-shell members of the
molecular density benchmark set of Ref.~\cite{Mezei2017}, assembled there to
provide chemically relevant densities against composite CCSDTQ references.
H$_2$ and Li$_2$ are omitted here because they have too few occupied orbitals to
support the fits of Sec.~\ref{sec:fits}. BH$_3$ is omitted because the free monomer
dimerizes to B$_2$H$_6$ and no reliable gas-phase VIE exists.
We call the molecules H$_2$O, CO, HF and N$_2$, whose ionization energies all come from Table~I of
Ref.~\cite{CGB2002}, the primary set, and it is used for every fit.

Inversions were performed with the inverse GKS method of Subramanian
\textit{et al.}~\cite{invGKS2026}, implemented in the \texttt{invDFT} real-space finite-element framework~\cite{invDFT,DFTFE2022}, whose methodological origin is
Ref.~\cite{Kanungo2019}. The local multiplicative remainder is obtained by a
constrained minimization in the manner of Wu and Yang~\cite{WuYang2003},
extended to the GKS operator through the explicit presence of the nonlocal
exchange term, whose action on the orbitals is evaluated by the adaptively
compressed exchange approximation~\cite{ACE2016}. Two numerical safeguards
matter here. A correction $\Delta\rho$ is added to the target density to remove the atomic-orbital basis-set error near the
nuclei~\cite{Kanungo2019}, and the known
$-(1-\alpha)/r$ leading decay is enforced on the multiplicative remainder in the
low-density region, $\rho < 10^{-6}$. Five exact-exchange fractions were used,
$\alpha = 0, 0.25, 0.5, 0.75, 1.0$.

\subsection{Experimental vertical ionization energies and their provenance}

Most values in Table~\ref{tab:orbcmp} are established data from
Ref.~\cite{CGB2002}, which draws on
Refs.~\cite{Baltzer1992,Siegbahn1969,Turner1971,Cavigliasso1999}, and
independently verified by Ref.~\cite{GBB2003}. The only
discrepancy among the twenty-one shared values is the HF~$2\sigma$ ionization
energy, for which Ref.~\cite{GBB2003} gives $39.7$~eV rather than the $39.60$~eV
used here. The remaining problematic entries stem from two distinct limitations.

\begingroup\small
\begin{longtable}{@{}llrrrrrl@{}}
\caption{Comparison of full-configuration-interaction vertical ionization energies (FCI/aug-cc-pVQZ, Ref.~\cite{MarieLoos2024}), the exact Kohn--Sham orbital energies of Ref.~\cite{GBB2003}, and the present inverse-GKS orbital energies at $\alpha=0$, with experimental vertical ionization energies. Orbital energies are given as $-\varepsilon_i$ (eV); the FCI column is an ionization energy, not an orbital energy. MAE denotes the mean absolute error against the experimental VIEs, given separately
  for the outer valence levels ($I < 21.2$~eV), MAE(val), the inner valence levels ($21.2 < I < 100$~eV), MAE(inner), and the core levels ($I > 100$~eV), MAE(core). Orbitals marked \emph{excl.} are excluded from all fits and averages. $^{\dagger}$Ref.~\cite{CGB2002}; $^{\ddagger}$Ref.~\cite{GBB2003}, the same construction in a different basis.}\label{tab:orbcmp}\\
\toprule
Molecule & MO & FCI/QZ & KS$^{\dagger}$ & KS$^{\ddagger}$ & invGKS & Expt. & \\
\midrule
\endfirsthead
\multicolumn{8}{@{}l}{\emph{Table \thetable\ continued}}\\
\toprule
Molecule & MO & FCI/QZ & KS$^{\dagger}$ & KS$^{\ddagger}$ & invGKS & Expt. & \\
\midrule
\endhead
\midrule
\multicolumn{8}{r@{}}{\emph{continued on next page}}\\
\endfoot
\bottomrule
\multicolumn{8}{@{}p{0.93\textwidth}@{}}{$^{a}$~Table I of Ref.~\cite{CGB2002} and references therein.}\\
\multicolumn{8}{@{}p{0.93\textwidth}@{}}{$^{b}$~Ref.~\cite{CGB2002} set, via Table S3 of the supplementary material of Ref.~\cite{Ranasinghe2019}.}\\
\multicolumn{8}{@{}p{0.93\textwidth}@{}}{$^{c}$~Not traced to a primary source; no vertical ionization energy is defined at this depth (Sec. 3). Excluded from all fits and averages.}\\
\multicolumn{8}{@{}p{0.93\textwidth}@{}}{$^{d}$~Ref.~\cite{Jolly1984}, from Ref.~\cite{Carroll1974}.}\\
\multicolumn{8}{@{}p{0.93\textwidth}@{}}{$^{e}$~Ref.~\cite{Berkowitz1979} via Ref.~\cite{MarieLoos2024}. The $1\pi$ entry is the degeneracy-weighted mean of the 11.50 and 11.67 eV spin-orbit components.}\\
\multicolumn{8}{@{}p{0.93\textwidth}@{}}{$^{f}$~Solid, gold-referenced XPS, Ref.~\cite{RoLinton1992}. Crystal binding energies, not molecular VIEs; excluded (Sec. 3).}\\
\multicolumn{8}{@{}p{0.93\textwidth}@{}}{$^{g}$~Fails the sign test of Sec. 3: the Li 1s of LiH must exceed the free-atom 64.4--64.9 eV. Excluded.}\\
\endlastfoot
H$_2$O  & $1b_1$ & 12.74 & 12.63 & 12.62 & 12.45 & 12.62$^{a}$ &  \\
        & $3a_1$ & 14.96 & 14.78 & 14.73 & 14.54 & 14.74$^{a}$ &  \\
        & $1b_2$ & 19.06 & 18.46 & 18.33 & 18.30 & 18.55$^{a}$ &  \\
        & $2a_1$ & --- & 30.89 & 30.72 & 30.63 & 32.20$^{a}$ &  \\
        & $1a_1$ & --- & 516.96 & 517.48 & 514.61 & 539.90$^{a}$ &  \\
\multicolumn{8}{@{}l}{\quad MAE (FCI): val 0.284} \\
\multicolumn{8}{@{}l}{\quad MAE (KS$^{\dagger}$): val 0.047, inner 1.310, core 22.940} \\
\multicolumn{8}{@{}l}{\quad MAE (KS$^{\ddagger}$): val 0.077, inner 1.480, core 22.420} \\
\multicolumn{8}{@{}l}{\quad MAE (invGKS): val 0.207, inner 1.571, core 25.286} \\
\addlinespace
CO      & $5\sigma$ & 13.97 & 14.01 & 14.01 & 14.04 & 14.01$^{a}$ &  \\
        & $1\pi$ & 17.02 & 16.77 & 16.80 & 16.93 & 16.90$^{a}$ &  \\
        & $1\pi$ & 17.02 & 16.77 & 16.80 & 16.93 & 16.90$^{a}$ &  \\
        & $4\sigma$ & 19.70 & 19.33 & 19.37 & 19.28 & 19.72$^{a}$ &  \\
        & $3\sigma$ & --- & 34.69 & 34.70 & 34.59 & 38.30$^{a}$ &  \\
        & $2\sigma$ & --- & 278.83 & 279.27 & 277.15 & 296.21$^{a}$ &  \\
        & $1\sigma$ & --- & 519.71 & 519.92 & 517.22 & 542.55$^{a}$ &  \\
\multicolumn{8}{@{}l}{\quad MAE (FCI): val 0.072} \\
\multicolumn{8}{@{}l}{\quad MAE (KS$^{\dagger}$): val 0.162, inner 3.610, core 20.110} \\
\multicolumn{8}{@{}l}{\quad MAE (KS$^{\ddagger}$): val 0.137, inner 3.600, core 19.785} \\
\multicolumn{8}{@{}l}{\quad MAE (invGKS): val 0.130, inner 3.707, core 22.193} \\
\addlinespace
HF      & $1\pi$ & 16.21 & 16.18 & 16.18 & 15.96 & 16.19$^{a}$ &  \\
        & $1\pi$ & 16.21 & 16.18 & 16.18 & 15.96 & 16.19$^{a}$ &  \\
        & $3\sigma$ & 20.10 & 19.92 & 19.90 & 19.69 & 19.90$^{a}$ &  \\
        & $2\sigma$ & --- & 36.76 & 36.77 & 36.50 & 39.60$^{a}$ &  \\
        & $1\sigma$ & --- & 668.46 & 668.33 & 664.81 & 694.23$^{a}$ &  \\
\multicolumn{8}{@{}l}{\quad MAE (FCI): val 0.083} \\
\multicolumn{8}{@{}l}{\quad MAE (KS$^{\dagger}$): val 0.013, inner 2.840, core 25.770} \\
\multicolumn{8}{@{}l}{\quad MAE (KS$^{\ddagger}$): val 0.007, inner 2.830, core 25.900} \\
\multicolumn{8}{@{}l}{\quad MAE (invGKS): val 0.226, inner 3.103, core 29.423} \\
\addlinespace
N$_2$   & $3\sigma_g$ & 15.54 & 15.57 & 15.57 & 15.49 & 15.58$^{a}$ &  \\
        & $1\pi_u$ & 16.98 & 16.68 & 16.68 & 16.75 & 16.93$^{a}$ &  \\
        & $1\pi_u$ & 16.98 & 16.68 & 16.68 & 16.75 & 16.93$^{a}$ &  \\
        & $2\sigma_u$ & 18.74 & 18.77 & 18.77 & 18.73 & 18.75$^{a}$ &  \\
        & $2\sigma_g$ & --- & 33.69 & 33.69 & 33.56 & 37.30$^{a}$ &  \\
        & $1\sigma_u$ & --- & 389.76 & 389.72 & 387.82 & 409.98$^{a}$ &  \\
        & $1\sigma_g$ & --- & 389.76 & 389.76 & 387.86 & 409.98$^{a}$ &  \\
\multicolumn{8}{@{}l}{\quad MAE (FCI): val 0.037} \\
\multicolumn{8}{@{}l}{\quad MAE (KS$^{\dagger}$): val 0.132, inner 3.610, core 20.220} \\
\multicolumn{8}{@{}l}{\quad MAE (KS$^{\ddagger}$): val 0.132, inner 3.610, core 20.240} \\
\multicolumn{8}{@{}l}{\quad MAE (invGKS): val 0.115, inner 3.737, core 22.138} \\
\addlinespace
F$_2$   & $1\pi_g$ & 15.81 & --- & --- & 15.93 & 15.87$^{b}$ &  \\
        & $1\pi_g$ & 15.81 & --- & --- & 15.93 & 15.87$^{b}$ &  \\
        & $1\pi_u$ & 18.94 & --- & --- & 19.31 & 18.80$^{b}$ &  \\
        & $1\pi_u$ & 18.94 & --- & --- & 19.31 & 18.80$^{b}$ &  \\
        & $3\sigma_g$ & 21.15 & --- & --- & 21.70 & 21.10$^{b}$ &  \\
        & $2\sigma_u$ & --- & --- & --- & 34.25 & 38.00$^{c}$ & excl. \\
        & $2\sigma_g$ & --- & --- & --- & 40.47 & 38.00$^{c}$ & excl. \\
        & $1\sigma_u$ & --- & --- & --- & 667.96 & 696.70$^{d}$ &  \\
        & $1\sigma_g$ & --- & --- & --- & 667.96 & 696.70$^{d}$ &  \\
\multicolumn{8}{@{}l}{\quad MAE (FCI): val 0.092} \\
\multicolumn{8}{@{}l}{\quad MAE (invGKS): val 0.345, core 28.742} \\
\addlinespace
LiF     & $1\pi$ & 11.54 & --- & --- & 11.66 & 11.59$^{e}$ &  \\
        & $1\pi$ & 11.54 & --- & --- & 11.66 & 11.59$^{e}$ &  \\
        & $4\sigma$ & 12.02 & --- & --- & 11.95 & 11.94$^{e}$ &  \\
        & $3\sigma$ & --- & --- & --- & 30.24 & 30.80$^{f}$ & excl. \\
        & $2\sigma$ & --- & --- & --- & 56.51 & 56.80$^{f}$ & excl. \\
        & $1\sigma$ & --- & --- & --- & 659.61 & 686.50$^{f}$ & excl. \\
\multicolumn{8}{@{}l}{\quad MAE (FCI): val 0.057} \\
\multicolumn{8}{@{}l}{\quad MAE (invGKS): val 0.052} \\
\addlinespace
LiH     & $2\sigma$ & --- & --- & --- & 8.18 & 7.90$^{a}$ &  \\
        & $1\sigma$ & --- & --- & --- & 56.46 & 60.90$^{g}$ & excl. \\
\multicolumn{8}{@{}l}{\quad MAE (invGKS): val 0.284} \\
\addlinespace
\end{longtable}
\endgroup

The first is the breakdown of the single-quasiparticle approximation, which
obscures well-defined vertical ionization peaks. For F$_2$, the literature
commonly assigns $38.0$~eV to both the $2\sigma_g$ and $2\sigma_u$ orbitals,
despite a calculated separation of $6.2$~eV, and no primary source for that value
exists; the F$_2$ $1s$ energy, by contrast, is traceable to the ESCA measurement
of Ref.~\cite{Carroll1974} through the compilation of Ref.~\cite{Jolly1984}, and
the outer-valence values to Ref.~\cite{Ranasinghe2019}. This is physically consistent with strong electron correlation
distributing transition intensity across multiple satellite states (secondary
peaks in the photoelectron spectrum that accompany a primary ionization and
carry part of its intensity) much as it
does for the analogous CO~$3\sigma$ ionization, where the spectroscopic
factor (the pole strength, that is, the squared norm of the Dyson orbital,
measuring the fraction of the total ionization intensity carried by the primary
transition) is only $0.344$ and three satellites carrying $0.177$,
$0.083$ and $0.171$ span $37.35$ to $40.35$~eV~\cite{GBB2003}. Following
Ref.~\cite{MarieLoos2024}, which reports three valence ionizations and one
satellite for F$_2$ and assigns no inner-valence value, the F$_2$ inner-valence
entries are included only for completeness and are excluded from all statistical
fits.

The second involves sample heterogeneity and phase disparities. Photoelectron
spectra of uncontrolled LiF vapor blend monomer and oligomer signals, LiF being
the extreme case of the alkali halide series for vapor
association~\cite{Eisenstadt1958}. Using
monomer-resolved data~\cite{Berkowitz1979}, the spin--orbit split $1\pi$
ionization yields a degeneracy-weighted mean of $11.585$~eV, so that $1\pi$ is
less tightly bound than $4\sigma$ ($11.94$~eV) by $0.36$~eV. Our inversion,
working from the electron density alone, separates them by $0.30$~eV in the same
order.

Finally, the LiF deep levels commonly cited ($686.5$, $56.8$ and $30.8$~eV)
derive from condensed-phase measurements referenced to a crystal Fermi
level~\cite{RoLinton1992}, and such values are invalid for isolated molecules.
The $60.9$~eV commonly cited for the Li~$1s$ level of LiH fails a sign test:
cationic lithium in molecular LiH must have
a $1s$ binding energy exceeding the free-atom value of
$64.4$--$64.9$~eV~\cite{Jolly1984}, a condition the cited number fails. All four
of these levels are excluded from the analysis.

\section{Results}\label{sec:res}

Two quantities are reported throughout. The first is the absolute deviation
$I_k+\varepsilon_k$, which enters Eq.~\eqref{eq:fit} directly but is contaminated
by the far-field alignment of the spectrum. The second is the HOMO-referenced
deviation $D_k$ of Eq.~\eqref{eq:Dk}, which is free of that contamination and
measures how well the GKS ladder reproduces the photoelectron ladder. Results
resting on the HOMO-referenced quantity are the more robust, and we say which is used
at each point.

\subsection{Validation for the Kohn--Sham system}\label{sec:valid}

Table~\ref{tab:orbcmp} sets the present orbital energies at $\alpha=0$, where
the GKS operator reduces to the Kohn--Sham one, against two earlier inverse-KS
constructions (those of CGB~\cite{CGB2002} and of GBB~\cite{GBB2003}), and
against the full-configuration-interaction vertical ionization energies of
Ref.~\cite{MarieLoos2024}. The three inversions are not equivalent
calculations. They differ in the target density, CGB and GBB using
configuration-interaction densities and the present work CCSD densities, and
they differ in basis. CGB and GBB also shift $\eHOMO$ explicitly to align with
$-\IHOMO$, as we do. The FCI column is of a different kind again: it is an
ionization energy computed directly, not an orbital energy. 
On the fourteen valence orbitals common to all four columns of Table~\ref{tab:orbcmp}, the mean absolute errors against experiment are 0.16 eV for the present work, 0.10 eV for both earlier inversions and 0.11 eV for the FCI energies.

To analyze different regions of the spectrum, we
divide the orbitals into three classes using the He(I) criterion of
Ref.~\cite{CGB2002}: core ($I > 100$~eV), inner valence ($21.2 < I < 100$~eV)
and outer valence ($I < 21.2$~eV). 
The comparison that matters for what follows is of the CGB coefficient itself.
Write $c_{\mathrm{core}}(\alpha)$ for the coefficient of Eq.~\eqref{eq:fit}
fitted to the core values of $I_k+\varepsilon_k$ alone; it is not the step-height coefficient $\Kg$ of Eq.~\eqref{eq:stepheight}. At $\alpha=0$ we
obtain $c_{\mathrm{core}}(0) = 1.132 \pm 0.028~\eVh$, against $1.028~\eVh$
extracted from Table~I of Ref.~\cite{CGB2002}. An independent atomic check gives
the same order: for Ne~$1s$, Table~I of Ref.~\cite{CGB2002} gives
$I+\varepsilon = 31.82$~eV against
$\sqrt{\varepsilon_N-\varepsilon_k} = 28.58~\eVh$, i.e.\
$c_{\mathrm{core}}(0) = 1.11~\eVh$.

The core orbital energies themselves are more method-sensitive than the valence
ones, and this bounds what can be concluded from the fitted coefficient. The two
inverse-KS columns are the same construction in two different basis sets and
differ by up to $0.52$~eV; our core energies are shallower than either by $1.7$
to $3.7$~eV, three to seven times that spread and one-signed. The abscissa is
unaffected (our $\sqrt{\eHOMO-\varepsilon_{1s}}$ values agree with those of
Ref.~\cite{CGB2002} to better than $0.08~\eVh$), so the discrepancy resides in
$I_k+\varepsilon_k$ at the core. Across genuinely different inversion methods the
spread is larger still: for atoms, the constructions of Zhao, Morrison and
Parr~\cite{ZMP1994} and of Ref.~\cite{CGB2002} differ by $7.6$~eV for the Be
$1s$ orbital energy~\cite{Shahi2024}. Section~\ref{sec:sys} quantifies what this
does to $\alpha^{*}$.

\subsection{The inverse-GKS spectrum}\label{sec:spectrum}

Table~\ref{tab:spectrum} reports the occupied spectra themselves. These are, to
our knowledge, the first inverse generalized Kohn--Sham calculations on
polyatomic molecules, and the first at any $\alpha>0$. The density is held at the
CCSD/aug-cc-pVQZ result for every $\alpha$, so what separates the columns in Table~\ref{tab:spectrum} 
is the one-particle operator, a separation no
self-consistent hybrid calculation can perform, since there the density and the
operator change together.

\begingroup\small
\begin{longtable}{@{}llrrrrr@{}}
\caption{Inverse generalized Kohn--Sham orbital energies $-\varepsilon_i$ (eV) for the seven molecules at fixed CCSD/aug-cc-pVQZ density. The exact-exchange fraction $\alpha$ parametrizes the one-particle operator only; the density is identical in every column. Each block closes with the deviation of the reference level (that of the first ionization) from $I_{\mathrm{HOMO}}$, that is $\varepsilon_{\mathrm{HOMO}}^{\alpha}+I_{\mathrm{HOMO}}$. Orbitals marked \emph{excl.} carry no reliable
  experimental counterpart and enter no fit; their eigenvalues are reported  for completeness.}\label{tab:spectrum}\\
\toprule
Molecule & MO & $\alpha=0$ & $0.25$ & $0.5$ & $0.75$ & $1.0$ \\
\midrule
\endfirsthead
\multicolumn{7}{@{}l}{\emph{Table \thetable\ continued}}\\
\toprule
Molecule & MO & $\alpha=0$ & $0.25$ & $0.5$ & $0.75$ & $1.0$ \\
\midrule
\endhead
\midrule
\multicolumn{7}{r@{}}{\emph{continued on next page}}\\
\endfoot
\bottomrule
\endlastfoot
H$_2$O  & $1b_1$ & 12.45 & 12.74 & 13.00 & 13.24 & 13.96 \\
        & $3a_1$ & 14.54 & 14.84 & 15.11 & 15.37 & 16.10 \\
        & $1b_2$ & 18.30 & 18.63 & 18.94 & 19.22 & 19.95 \\
        & $2a_1$ & 30.63 & 32.10 & 33.55 & 34.98 & 36.88 \\
        & $1a_1$ & 514.61 & 525.73 & 536.82 & 547.96 & 559.62 \\
\multicolumn{2}{@{}l}{\quad HOMO deviation} & +0.17 & -0.12 & -0.38 & -0.62 & -1.34 \\
\addlinespace
CO      & $5\sigma$ & 14.04 & 14.19 & 14.69 & 15.04 & 15.76 \\
        & $1\pi$ & 16.93 & 16.92 & 17.27 & 17.45 & 17.99 \\
        & $1\pi$ & 16.93 & 16.92 & 17.27 & 17.45 & 17.99 \\
        & $4\sigma$ & 19.28 & 19.74 & 20.56 & 21.23 & 22.27 \\
        & $3\sigma$ & 34.59 & 36.03 & 37.82 & 39.46 & 41.46 \\
        & $2\sigma$ & 277.15 & 284.90 & 293.07 & 300.87 & 309.24 \\
        & $1\sigma$ & 517.22 & 528.22 & 539.52 & 550.85 & 562.43 \\
\multicolumn{2}{@{}l}{\quad HOMO deviation} & -0.03 & -0.18 & -0.68 & -1.03 & -1.75 \\
\addlinespace
HF      & $1\pi$ & 15.96 & 16.30 & 16.52 & 16.86 & 17.49 \\
        & $1\pi$ & 15.96 & 16.30 & 16.52 & 16.86 & 17.49 \\
        & $3\sigma$ & 19.69 & 19.99 & 20.16 & 20.45 & 21.01 \\
        & $2\sigma$ & 36.50 & 38.13 & 39.64 & 41.28 & 43.19 \\
        & $1\sigma$ & 664.81 & 677.26 & 689.60 & 702.18 & 714.84 \\
\multicolumn{2}{@{}l}{\quad HOMO deviation} & +0.24 & -0.11 & -0.33 & -0.67 & -1.30 \\
\addlinespace
N$_2$   & $3\sigma_g$ & 15.49 & 15.74 & 16.25 & 16.69 & 17.52 \\
        & $1\pi_u$ & 16.75 & 16.69 & 16.87 & 16.96 & 17.44 \\
        & $1\pi_u$ & 16.75 & 16.69 & 16.87 & 16.96 & 17.44 \\
        & $2\sigma_u$ & 18.73 & 19.24 & 20.01 & 20.70 & 21.77 \\
        & $2\sigma_g$ & 33.56 & 34.94 & 36.57 & 38.12 & 40.06 \\
        & $1\sigma_u$ & 387.82 & 397.29 & 407.01 & 416.65 & 426.70 \\
        & $1\sigma_g$ & 387.86 & 397.34 & 407.07 & 416.73 & 426.80 \\
\multicolumn{2}{@{}l}{\quad HOMO deviation} & +0.09 & -0.16 & -0.67 & -1.11 & -1.94 \\
\addlinespace
F$_2$   & $1\pi_g$ & 15.93 & 15.98 & 16.36 & 17.02 & 17.66 \\
        & $1\pi_g$ & 15.93 & 15.98 & 16.36 & 17.02 & 17.65 \\
        & $1\pi_u$ & 19.31 & 19.52 & 20.06 & 20.87 & 21.66 \\
        & $1\pi_u$ & 19.31 & 19.52 & 20.06 & 20.87 & 21.66 \\
        & $3\sigma_g$ & 21.70 & 21.09 & 20.80 & 20.79 & 20.73 \\
        & $2\sigma_u$\ \emph{excl.} & 34.25 & 35.37 & 36.82 & 38.56 & 40.26 \\
        & $2\sigma_g$\ \emph{excl.} & 40.47 & 41.89 & 43.64 & 45.66 & 47.66 \\
        & $1\sigma_u$ & 667.96 & 680.13 & 692.65 & 705.49 & 718.27 \\
        & $1\sigma_g$ & 667.96 & 680.14 & 692.66 & 705.50 & 718.28 \\
\multicolumn{2}{@{}l}{\quad HOMO deviation} & -0.06 & -0.11 & -0.49 & -1.15 & -1.79 \\
\addlinespace
LiF     & $1\pi$ & 11.66 & 11.79 & 12.00 & 12.11 & 12.78 \\
        & $1\pi$ & 11.66 & 11.79 & 12.00 & 12.11 & 12.77 \\
        & $4\sigma$ & 11.95 & 12.17 & 12.47 & 12.66 & 13.40 \\
        & $3\sigma$\ \emph{excl.} & 30.24 & 31.76 & 33.37 & 34.87 & 36.93 \\
        & $2\sigma$\ \emph{excl.} & 56.51 & 58.17 & 60.75 & 63.15 & 66.17 \\
        & $1\sigma$\ \emph{excl.} & 659.61 & 671.95 & 684.44 & 696.83 & 709.71 \\
\multicolumn{2}{@{}l}{\quad HOMO deviation} & -0.07 & -0.20 & -0.41 & -0.53 & -1.19 \\
\addlinespace
LiH     & $2\sigma$ & 8.18 & 8.21 & 8.23 & 8.26 & 8.08 \\
        & $1\sigma$\ \emph{excl.} & 56.46 & 58.86 & 61.43 & 64.04 & 66.42 \\
\multicolumn{2}{@{}l}{\quad HOMO deviation} & -0.28 & -0.30 & -0.33 & -0.36 & -0.18 \\
\addlinespace
\multicolumn{7}{@{}l}{Error floor at $\alpha=0$: RMS $0.16$~eV, mean $+0.01$, SD $0.16$} \\
\end{longtable}
\endgroup

The spectrum stretches as $\alpha$ grows, and it stretches unevenly. Across the full range of admixtures
($0 \le \alpha \le 1$), the H$_2$O $1b_1$ energy deepens by $1.51$~eV while
$1a_1$ deepens by $45.0$~eV, so the core travels some thirty times further than
the outermost valence orbital. That disparity is why $\sqrt{\eHOMO-\varepsilon_k}$ spans $16$--$26~\eVh$ for
the $1s$ orbitals against $0$--$3~\eVh$ for the outer valence, and hence why the
core dominates the regressions of Sec.~\ref{sec:fits}.

The HOMO deviation ($\eHOMO^\alpha + I_{\mathrm{HOMO}}$) is a useful metric, 
as the theorem of Levy, Perdew and Sahni~\cite{PPLB1982,LPS1984,AvB1985} mandates it to be zero. In an
inverse calculation $\eHOMO$ is sensitive to the far-field decay of the target
density, and Gaussian densities lack the expected exponential decay, so
$\eHOMO$ carries an error induced by the incorrect far-field asymptotics of the
density. At $\alpha=0$ that error is $0.16$~eV root mean square over the seven
molecules, with a mean of $+0.01$~eV and a standard deviation of $0.16$~eV: the
inversion is unbiased at the HOMO and the residual is scatter. The HOMO
deviation therefore sets a floor beneath the other errors reported below.

The above floor is an artifact of the far-field asymptotics rather than of the
correlation treatment and is confirmed by the densities themselves.
Reference~\cite{Mezei2017} reports normed integral absolute deviations of the
CCSD density from composite CCSDTQ references for these same molecules, ranging
from $0.0004$ for LiH to $0.0032$ for CO. These do not track the HOMO deviations:
CO has the largest density error and among the smallest HOMO deviations, LiH the
smallest density error and the largest, and the correlation coefficient over the
seven molecules is $-0.53$. The quality of the correlated density is therefore
not what limits $\eHOMO$. The HOMO deviation is attributable to  the
one-particle basis rather than to the density. We note as a limitation that CCSD
densities are, on the same measure, an order of magnitude further from the
CCSDTQ reference than CCSDT densities are, and that inversions of CCSDT densities
have not been attempted here.

Broadly, the deviation increases with $\alpha$, reaching between $-1.3$ and
$-1.9$~eV for the primary set. This follows from the way the far-field decay of
the local multiplicative remainder is enforced (Sec.~\ref{sec:comp}). We impose a
$-(1-\alpha)/r$ decay in the low-density region, which is the correct leading
order for $\alpha<1$. As $\alpha$ tends to $1$, however, that term goes to zero
and terms decaying faster than $1/r$ dominate. The form of those terms is not
known a priori, so no appropriate far-field decay can be enforced on the local
multiplicative potential as $\alpha$ tends to $1$. Only LiH departs from this trend, with $-0.36$~eV at $\alpha=0.75$ and
$-0.18$~eV at $\alpha=1$, a difference within the scatter of the $\alpha>0$
inversions. To remove this numerical error, we align $\eHOMO^{\alpha}$ with $-I_{\mathrm{HOMO}}$ (Eq.~\eqref{eq:Dk}).

Symmetry-required degeneracies are preserved to within $0.003$~eV at every
$\alpha$ (Table~S2). Where levels cross, orbitals are assigned by symmetry
rather than by energy order: the $3\sigma_g$ level of F$_2$ rises above the
$1\pi_u$ pair near $\alpha=0.72$, and that of N$_2$ falls below it near
$\alpha=0.95$ (Table~\ref{tab:spectrum}).

\subsection{The CGB relation for GKS systems}\label{sec:fits}

Using the spectra of Table~\ref{tab:spectrum} we obtain the proportionality coefficient $c(\alpha)$ of Eq.~\eqref{eq:fit}
(Table~\ref{tab:fits}). The CGB relation holds at every
$\alpha$ considered here: the square-root form describes the occupied spectrum
with $R^{2} = 0.98$--$0.996$ at $\alpha=0$ and $0.98$--$0.99$ at $\alpha=1$. At
intermediate admixtures $R^{2}$ is lower, $0.84$ to $0.98$ at $\alpha=0.5$,
because $c(\alpha)$ passes through zero near $\alpha^{*}$, where the
square-root term has little left to describe. It is therefore not specific to the multiplicative Kohn--Sham system, which is
the only case for which it was proposed, but holds for GKS systems as well.

\begin{table}[htbp]\centering\small
\caption{The fitted coefficient $c$ of Eq.~(\ref{eq:fit}) in $\eVh$, with the $R^{2}$ of each fit on the line beneath, for the full occupied spectrum and for the outer-valence orbitals alone ($I<21.2$~eV). $s$ is the slope of the linear regression of $c$ on $\alpha$ and $\alpha^{*}$ its zero; that regression has $R^{2}\ge0.9996$ for every molecule. The F$_2$ row excludes the $2\sigma$ pair. LiF and LiH are omitted for the reasons given in Sec.~\ref{sec:fits}; in brief, LiH has two occupied
  orbitals, and LiF, with its deep levels excluded as crystal binding energies, has two distinct abscissae for a two-parameter fit. For N$_2$ at $\alpha=1$ see Sec.~\ref{sec:cgb}: the $1\pi_u$ level lies above the reference and takes abscissa zero. The closing block substitutes alternative reference standards for the experimental ionization energies and reports the resulting $\alpha^{*}$ and the outer-valence RMS and standard deviation. Only the Ranasinghe set~\cite{Ranasinghe2019} replaces core values; the others substitute valence orbitals only, so each substitution is partial; the count in parentheses is the number of primary-set orbitals it replaces.}
\label{tab:fits}
\setlength{\tabcolsep}{5pt}
\begin{tabular}{@{}lrrrrrrr@{}}
\toprule
 & $\alpha=0$ & $0.25$ & $0.5$ & $0.75$ & $1.0$ & $s$ & $\alpha^{*}$ \\
\midrule
\multicolumn{8}{@{}l}{\emph{All orbitals}}\\
H$_2$O & 1.189 & 0.672 & 0.166 & $-0.333$ & $-0.824$ & 2.012 & 0.587 \\
\quad $R^{2}$ & 0.984 & 0.975 & 0.843 & 0.974 & 0.992 & & \\
CO & 1.219 & 0.697 & 0.186 & $-0.313$ & $-0.801$ & 2.020 & 0.598 \\
\quad $R^{2}$ & 0.993 & 0.991 & 0.930 & 0.946 & 0.983 & & \\
HF & 1.176 & 0.682 & 0.196 & $-0.285$ & $-0.751$ & 1.929 & 0.606 \\
\quad $R^{2}$ & 0.991 & 0.990 & 0.984 & 0.992 & 0.991 & & \\
N$_2$ & 1.201 & 0.680 & 0.175 & $-0.315$ & $-0.784$ & 1.987 & 0.596 \\
\quad $R^{2}$ & 0.996 & 0.993 & 0.892 & 0.944 & 0.985 & & \\
F$_2$ & 1.187 & 0.683 & 0.187 & $-0.301$ & $-0.779$ & 1.966 & 0.599 \\
\quad $R^{2}$ & 0.991 & 0.991 & 0.939 & 0.953 & 0.984 & & \\
\addlinespace
\multicolumn{6}{@{}l}{\quad Primary set; intercept $c(0)=1.185\pm0.017$} & $1.987\pm0.041$ & $0.597\pm0.008$ \\
\multicolumn{6}{@{}l}{\quad Primary set $+$ F$_2$} & & $0.597\pm0.007$ \\
\multicolumn{6}{@{}l}{\quad Electron-gas coefficient $\Kg$, Eq.~(\ref{eq:stepheight})} & $1.9932$ & \\
\addlinespace
\multicolumn{8}{@{}l}{\emph{Outer valence only}}\\
H$_2$O & 0.033 & 0.014 & $-0.003$ & $-0.018$ & $-0.026$ & & \\
\quad $R^{2}$ & 0.962 & 0.996 & 0.408 & 0.841 & 0.929 & & \\
CO & 0.138 & 0.076 & $-0.016$ & $-0.119$ & $-0.239$ & & \\
\quad $R^{2}$ & 0.341 & 0.897 & 0.005 & 0.071 & 0.131 & & \\
HF & $-0.014$ & 0.010 & 0.038 & 0.065 & 0.103 & & \\
\quad $R^{2}$ & 1.000 & 1.000 & 1.000 & 1.000 & 1.000 & & \\
N$_2$ & $-0.024$ & $-0.164$ & $-0.394$ & $-0.696$ & $-0.989$ & & \\
\quad $R^{2}$ & 0.054 & 0.125 & 0.243 & 0.418 & 0.695 & & \\
\bottomrule
\end{tabular}
\par\medskip
\begin{minipage}{\textwidth}\small
  \emph{Sensitivity to the reference standard.} Refitting with alternative reference ionization energies in place of those of Table~\ref{tab:orbcmp} moves $\alpha^{*}$ by at most 0.003. For the primary set, each reference ionization energy set replaces the values used here for different numbers of orbitals, leading to slightly different values for $\alpha^{*}$. In the following, we report five other reference sets, the number of orbitals replaced ($n_r$), and the refitted value of $\alpha^{*}$:
  (i) Ref.~\cite{GBB2003}: $n_r=3$, $\alpha^{*} = 0.597 \pm 0.008$; (ii) Ref.~\cite{MarieLoos2024} (experimental values): $n_r=8$, $\alpha^{*}=0.596 \pm 0.008$;  (iii) Ref.~\cite{Ranasinghe2019}: $n_r=8$,  $\alpha^{*} = 0.594 \pm 0.009$; (iv) Ref.~\cite{MarieLoos2024} (FCI/aug-cc-pVTZ): $n_r=14$, $\alpha^{*}=0.596 \pm 0.010$; and (v) Ref.~\cite{MarieLoos2024} (FCI/aug-cc-pVQZ): $n_r=14$, $\alpha^{*}=0.595 \pm 0.010$. The corresponding outer-valence RMS and standard deviation, in eV, are 0.157/0.142, 0.178/0.141, 0.157/0.142, 0.251/0.163 and 0.247/0.163. Only the Ranasinghe set replaces core values, so every substitution is partial.
\end{minipage}
\end{table}

We find $c(\alpha)$ to be linear in the exact-exchange fraction,
\begin{equation}
c(\alpha) = c(0) - s\,\alpha ,
\qquad
c(0) = 1.185 \pm 0.017~\eVh ,
\qquad
s = 1.987 \pm 0.041~\eVh ,
\label{eq:linear}
\end{equation}
with $R^{2}_{\mathrm{lin}} \ge 0.9996$ for each molecule of the primary set. The
linearity is not obvious: the inversion depends nonlinearly on
$\hat{\Sigma}_{\mathrm{x}}$, which is built from the occupied orbitals of
$\hat{h}^{\alpha}$ itself.

The coefficient changes sign between $\alpha=0$ and $\alpha=1$, vanishing at
$\alpha^{*} = 0.597 \pm 0.008$ across the primary set; the individual values are
the last column of Table~\ref{tab:fits}, and adding F$_2$ leaves the mean at
$0.597 \pm 0.007$. LiF and LiH do not appear: LiH has two occupied orbitals, and
LiF, once its deep levels are excluded as crystal binding energies
(Sec.~\ref{sec:comp}), retains only three valence orbitals, two of them the degenerate $1\pi$ pair, so its fit has two distinct abscissae for two parameters and is exact.
Empirically, then, $\alpha \approx 0.6$ is the fraction of exact exchange at
which the HOMO-referenced GKS occupied spectrum corresponds to the negative of
the measured ionization energies, $\varepsilon_k^{0.6} \approx -I_k$.

Independent support comes from a functional parameterized on the same
information. QTP17 is a global hybrid with $\alpha = 0.62$ exact
exchange~\cite{Haiduke2018b}, obtained by imposing Bartlett's IP-eigenvalue
theorem with core states weighted most heavily. The agreement is not accidental:
both determinations are anchored on the core ionization energies, QTP17 through the weighting of its fit and $\alpha^{*}$ through the dominance of
the core orbitals in the regression, shown below. The QTP hybrid functionals are built on an analogy with
coupled-cluster theory; although both are correlated orbital theories, they are
different ones that can have different features, so the agreement is not a
derivation. QTP17 also inherits the penalty our analysis predicts: benchmarked
against $401$ valence ionizations of $63$ molecules, it shows a larger mean
signed deviation for the HOMOs ($-0.78$~eV) than for all valence states taken
together ($-0.18$~eV), in contrast to every other functional
tested~\cite{Ranasinghe2019}.

The numerical values of $c(0)$, $s$ and $\alpha^{*}$ are dominated by the core
orbitals, as they were in the original CGB work~\cite{CGB2002}. 
 
The dominance of the core in the regression
follows from the abscissa: $\sqrt{\eHOMO-\varepsilon_k}$ is $16$--$26~\eVh$ for
the $1s$ orbitals but only $0$--$3~\eVh$ for the outer valence. Consistently,
fitting the core orbitals together with the HOMO reproduces $\alpha^{*}$ to
within $0.01$ for every molecule. In form the square-root relation holds for all
orbitals; in magnitude it holds primarily at the core. Each inner-valence deviation falls linearly with $\alpha$, but at only
$0.53$--$0.59$ times the rate that Eq.~\eqref{eq:fit} predicts for its
abscissa; the six core orbitals fall at $0.93$--$0.99$ times that rate.

\begin{figure}[htbp]\centering
\includegraphics[width=\textwidth]{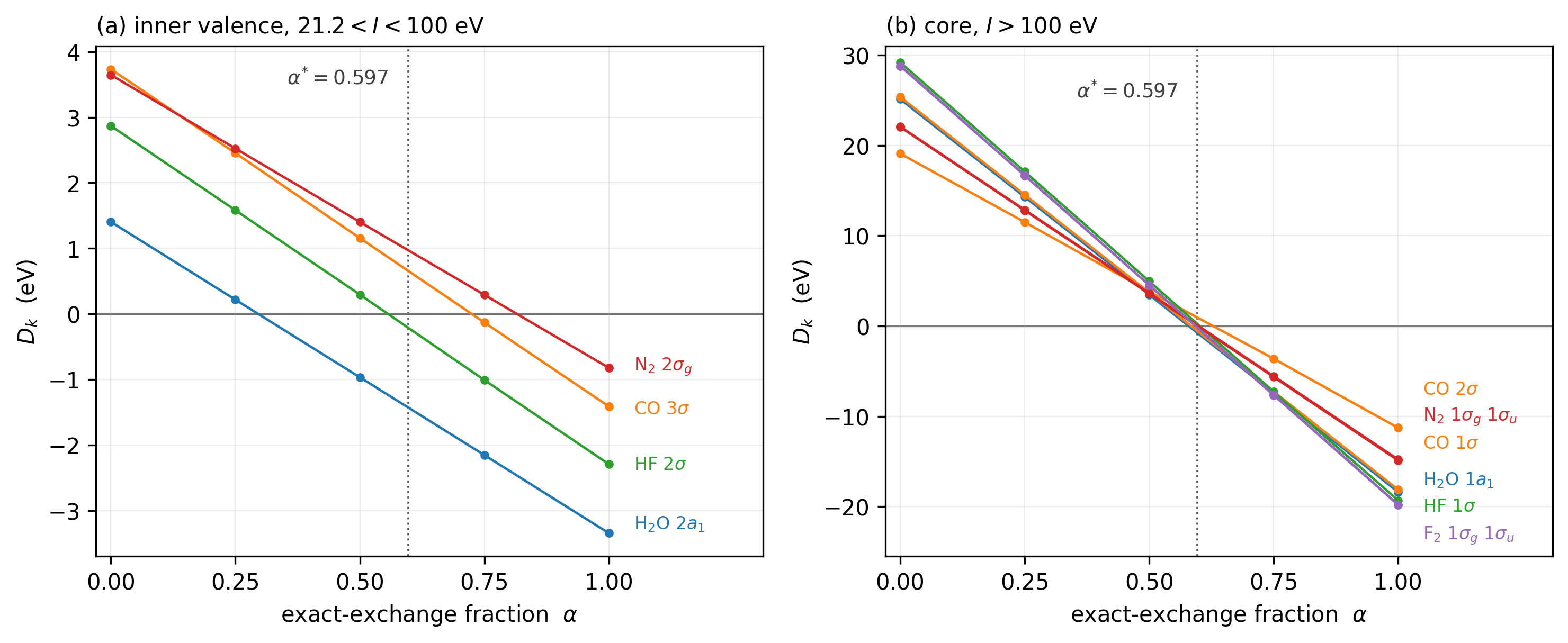}
\caption{HOMO-referenced deviation $D_k$ of Eq.~\eqref{eq:Dk} against
exact-exchange fraction $\alpha$. Every occupied orbital of the class is drawn
and labeled at the right of its $\alpha=1$ endpoint; orbitals of one molecule
that end at the same energy are written on one line, since they are degenerate.
(a) inner valence, $21.2 < I < 100$~eV, one orbital per molecule. (b) core,
$I>100$~eV, where CO, N$_2$ and F$_2$ each contribute two orbitals. The vertical
scales differ by a factor of eight and the two classes behave differently on
them: the inner-valence lines descend at nearly equal rates from unequal
starting points, so they cross zero at four separate admixtures spread over $0.3$ to $0.8$, whereas the core lines start spread over $10$~eV, converge as
they descend, and change sign together at $\alpha^{*}$. Orbitals excluded from
the fits (Table~\ref{tab:orbcmp}) are omitted, which leaves LiH and LiF out of both panels and F$_2$ out of panel (a). The outer-valence deviations, which start
near zero and fan outward without crossing, are shown in Fig.~S1.}
\label{fig:Dk}
\end{figure}

Table~\ref{tab:class} gives the mean and the standard deviation of $D_k$ for
each class. As functions of $\alpha$, the mean of $D_k$ changes sign for the
core, from $+23.8$ to $-16.1$~eV, and for the inner valence, from $+2.9$ to
$-2.0$~eV, whereas for the outer valence it stays positive, drifting only from
$+0.065$ to $+0.240$~eV; Figure~\ref{fig:Dk} shows the same behavior orbital by
orbital. The core and the inner valence thus have a zero crossing in the mean
and the outer valence does not, which is a further reason why $\alpha^{*}$ is
set by the core. In the outer valence the standard deviation grows from $0.14$
to $0.79$~eV: admixing exact exchange does not
systematically shift the outer-valence ladder of orbital energies but scatters
it.

\begin{table}[htbp]\centering\small
\caption{Deviation $D_k$ of Eq.~(\ref{eq:Dk}) from zero, by orbital class, in eV, for the primary set, with the HOMO excluded ($D_k=0$ by construction). Each block gives the mean $D_k$, its standard deviation and their resultant, $\mathrm{RMS}^2=\mathrm{mean}^2+\mathrm{SD}^2$; the minimum of each RMS row is in bold. $n$ counts the orbitals of each class entering the fits. The F$_2$ $2\sigma$ pair and the LiF and LiH deep levels are excluded throughout.}
\label{tab:class}
\begin{tabular}{@{}lrrrrrr@{}}
\toprule
 & $n$ & $\alpha=0$ & $0.25$ & $0.5$ & $0.75$ & $1.0$ \\
\midrule
Outer valence ($I<21.2$~eV) & 10 & 0.065 & 0.103 & 0.139 & 0.186 & 0.240 \\
\quad SD & & 0.143 & 0.202 & 0.380 & 0.582 & 0.791 \\
\quad RMS & & \textbf{0.157} & 0.227 & 0.404 & 0.611 & 0.826 \\
\addlinespace[2pt]
Inner valence ($21.2$--$100$~eV) & 4 & 2.912 & 1.694 & 0.471 & -0.751 & -1.968 \\
\quad SD & & 0.934 & 0.930 & 0.927 & 0.936 & 0.952 \\
\quad RMS & & 3.059 & 1.932 & \textbf{1.040} & 1.200 & 2.186 \\
\addlinespace[2pt]
Core ($I>100$~eV) & 6 & 23.806 & 13.840 & 3.861 & -6.137 & -16.129 \\
\quad SD & & 3.203 & 1.767 & 0.504 & 1.364 & 2.772 \\
\quad RMS & & 24.021 & 13.952 & \textbf{3.894} & 6.287 & 16.365 \\
\addlinespace[2pt]
\bottomrule
\end{tabular}
\end{table}

It follows that $\alpha^{*}$ is a statement about the core, not about the
valence spectrum. In particular, $\alpha^{*} \approx 0.6$ is not the admixture
minimizing the outer-valence error, which for the primary set is $\alpha=0$
(Table~\ref{tab:class}); N$_2$ and F$_2$ carry most of the outer-valence
degradation (Fig.~S1). Refitting the outer valence alone, the slope at
$\alpha=0$ lies between $-0.024$ and $0.138~\eVh$, indistinguishable from zero:
no significant square-root relation remains once the core orbitals are removed.
HF is the exception: its outer valence has two distinct abscissae, the
degenerate $1\pi$ pair and $3\sigma$, so its refit is exact and the $R^{2}$ of
$1.000$ carries no information.

The closing block of Table~\ref{tab:fits} tests these conclusions against the
choice of experimental reference, and is discussed in Sec.~\ref{sec:sys}.

\section{Discussion}\label{sec:disc}

The methodological point of this work is a separation that self-consistent
calculations cannot perform. In a self-consistent hybrid, changing $\alpha$
changes both the density and the operator, and the resulting change in
eigenvalues mixes the two. Here the density is fixed at the near-exact result, so
the entire $\alpha$-dependence reported above is attributable to the operator.

The consequence is a statement of what exact-exchange admixture does to occupied eigenvalues. With
the correct density supplied, adding exact exchange does not make the occupied eigenvalues closer to the VIEs. 
It deepens them, and in terms of HOMO-referenced deviations it either
leaves the valence orbitals unchanged (H$_2$O, HF) or degrades them (N$_2$, F$_2$, CO); in no
molecule does it improve them beyond the $0.16$~eV error floor. Self-consistent
hybrids are a different matter, since there the density changes as well:
B3LYP occupied orbital energies deviate from the measured ionization energies
by $3.05$~eV on average over $56$ orbitals of $12$ molecules~\cite{Yang2020},
against $0.19$~eV over all our valence orbitals at $\alpha=0$.

The results presented here are valid for weakly correlated systems.
Reference~\cite{invGKS2026} shows that the common approximation
$\delta \vxc^{\alpha} \approx (1-\alpha)v_{\mathrm{x}}^{\mathrm{KS}}
+ v_{\mathrm{c}}^{\mathrm{KS}}$ holds for such systems but fails for stretched
ones, the mean-field energies differing by more than $11$~mHa for LiH and CO at
extended bond lengths. Whether the $\alpha$-dependence reported here is valid
for strongly correlated systems is therefore an open question.

\subsection{The measured slope and the electron-gas step-height coefficient}

Equation~(4.4) of Ref.~\cite{GBB2003} splits $\varepsilon_k + I_k$ into a
response term minus a spectroscopic term. Reference~\cite{GBB2003} treats only
the Kohn--Sham system ($\alpha=0$); we take the same split to hold for the GKS systems. Dividing each term by
$\sqrt{\eHOMO-\varepsilon_k}$ and averaging over the core orbitals defines two
coefficients, $K_{\mathrm{resp}}$ and $K_{\mathrm{spec}}$, and since
$\varepsilon_k + I_k$ is their difference, the CGB coefficient fitted to the
core is $c_{\mathrm{core}}(0) = K_{\mathrm{resp}} - K_{\mathrm{spec}}$. If the
spectroscopic part does not vary with $\alpha$, the whole of $dc/d\alpha$ comes
from the response part, and a response matrix element divided by the abscissa
should be comparable with $s$.

Reference~\cite{GBB2003} tabulates the response matrix elements
$(\mathbf{M}^{-1}\bm{\varepsilon}^{\mathrm{resp}})_i$, which makes
$K_{\mathrm{resp}}$ measurable:
$K_{\mathrm{resp}} = 1.972 \pm 0.040~\eVh$ over the six core orbitals of CO,
HF, H$_2$O and HCN (Table~S3). $K_{\mathrm{resp}}$ is thus the counterpart of $\Kg$
for the exact Kohn--Sham system, measured rather than modelled. $K_{\mathrm{resp}}$ and our slope
$s = 1.987 \pm 0.041~\eVh$ both agree with the electron-gas value $\Kg$ of
Eq.~\eqref{eq:stepheight}. The spectroscopic
data of Ref.~\cite{GBB2003} give $K_{\mathrm{spec}} = 0.959~\eVh$, so
$c_{\mathrm{core}}(0) = 1.972 - 0.959 = 1.013~\eVh$, matching the
$1.028~\eVh$ obtained from Table~I of Ref.~\cite{CGB2002}. The CGB coefficient
is thus the difference of two comparable terms, neither of them small: the
response part alone is nearly twice the coefficient it produces. The relation
$\varepsilon_k \approx -I_k$ holds by cancellation, as Ref.~\cite{GBB2003}
emphasizes, which is why changing the operator at fixed density can disturb it.

If the spectroscopic part is independent of $\alpha$, the response term in
$c(\alpha)$ falls linearly from about $\Kg$ at $\alpha=0$ to about zero at
$\alpha=1$, where exchange is carried entirely by the nonlocal operator. We
therefore conjecture that exact-exchange admixture removes the response
contribution at the electron-gas rate,
\begin{equation}
\frac{dc}{d\alpha} = -\Kg ,
\label{eq:conj}
\end{equation}
while the spectroscopic contribution does not vary with $\alpha$. The linearity
of Eq.~\eqref{eq:linear} would then be a consequence rather than an
observation. The response matrix elements of Ref.~\cite{GBB2003} are
$\alpha=0$ quantities containing both the exchange and the correlation
response, so the agreement of $K_{\mathrm{resp}}$ with $s$ is consistent with a
small correlation response. We present this as a conjecture, not a result. The coefficient $c$
is a slope of $\varepsilon_k + I_k$, whereas $\Kg$ is the coefficient of the
step height $w_i$; Refs.~\cite{CGB2002,GLLB1995,GBB2003} connect the two, but
they are not the same quantity, and part of the agreement may be fortuitous.
The identification also rests on the Kohn--Sham system, and no analogue has
been derived for the GKS operator at $\alpha>0$. The conjecture makes a testable
prediction: since $\Kg$ is a universal constant, $dc/d\alpha$ should be independent of
the molecule, and the observed spread of $1.929$--$2.020~\eVh$ should contract as
the densities improve and the inversions are converged more tightly.

\subsection{Systematic uncertainty in $\alpha^{*}$}\label{sec:sys}

$\alpha^{*} = c(0)/s$, and $c(0)$ is determined predominantly by the core
orbitals (Sec.~\ref{sec:fits}). The core discrepancy of Sec.~\ref{sec:valid}
therefore propagates into $\alpha^{*}$ with a known sign: our core orbital
energies are shallower than those of Ref.~\cite{CGB2002}, which increases
$I_k+\varepsilon_k$ at the core and hence $c(0)$. To estimate the uncertainty in $\alpha^{*}$, we
substitute the $\alpha=0$ eigenvalues of Ref.~\cite{CGB2002} for our own. This lowers the mean fitted
coefficient at $\alpha=0$ from $1.197$ (Table~\ref{tab:fits}) to $1.078~\eVh$.
This should not be confused with $c(0) = 1.185~\eVh$ of
Eq.~\eqref{eq:linear}, which is the intercept of the linear fit of $c(\alpha)$
over all five admixtures. If $s$ is unchanged, the shift of $-0.119~\eVh$ lowers
$\alpha^{*}$ by $0.119/s = 0.060$, to $0.537$. Substituting the eigenvalues of an independent inversion, from a different density and basis, thus lowers $\alpha^{*}$ by $0.06$. The linearity of $c(\alpha)$, the ordering of the
orbital classes and the agreement of $s$ with $\Kg$ do not depend on $c(0)$.

The choice of experimental reference matters much less to the value of
$\alpha^{*}$. We substituted five alternative reference sets, drawn from three
sources, for our reference values: (i) the compilation of
Ref.~\cite{GBB2003}; (ii) the experimental values quoted by
Ref.~\cite{MarieLoos2024}, which for H$_2$O are Ning's electron-momentum
determinations, $0.08$ and $0.17$~eV above Chong's for $3a_1$ and $1b_2$;
(iii) the benchmark set of Ref.~\cite{Ranasinghe2019}, whose core values lie
between $0.02$~eV above and $0.31$~eV below ours; and (iv, v) the
full-configuration-interaction values of Ref.~\cite{MarieLoos2024} with the
aug-cc-pVTZ and aug-cc-pVQZ basis sets. Across all five, $\alpha^{*}$ moves by
at most $0.003$ (Table~\ref{tab:fits}). The largest shift comes from the
benchmark set of Ref.~\cite{Ranasinghe2019}, the only one of the five that
changes the core values.

\section{Outlook: the copper oxide anions}

The systems studied here are normally correlated closed-shell $sp$ molecules, for
which the CGB relation was expected to hold and does. A stronger test is a system
in which it is expected to fail.

The measured photoemission spectra of the anions Cu$_2$O$^-$, CuO$^-$,
CuO$_2^-$ and CuO$_3^-$ are reproduced by no density functional approximation yet
tested~\cite{Shahi2024}. LSDA, its Perdew--Zunger and locally scaled
self-interaction corrections, global and range-separated hybrids, and the QTP02
correlated-orbital-theory functional all fail, in mutually inconsistent ways;
QTP02, which achieves a mean absolute deviation of $0.36$~eV for the highest
occupied orbital energies of small organic molecules~\cite{Haiduke2018}, gives a
highest occupied orbital energy in CuO$^-$ and CuO$_3^-$ that appears far too
negative. The failure has been attributed variously to self-interaction error and
to strong correlation, and the two have not been disentangled.

Inverse GKS separates the two contributions for the same reason it does so here:
it eliminates the density as a variable. Given an accurate correlated density for
CuO$^-$ ($38$ electrons), inversion yields the almost-exact KS or GKS potential
and orbital energies, and the residual discrepancy with the measured spectrum is
then unambiguously not a density error. This is the experiment proposed in the
concluding paragraph of Ref.~\cite{Shahi2024}, which points to the inversion
methodology of Ref.~\cite{Kanungo2024}. The present work supplies the missing
precondition: a demonstration on systems where the answer is known that the
inversion reproduces valence eigenvalues to better than $0.2$~eV, together with a
quantitative bound on the corresponding core-orbital error.

Two caveats should be stated in advance. An open-shell transition-metal anion
with $3d$ electrons is a considerably harder inversion target than a closed-shell
$sp$ molecule, and the requisite reference density is itself a research problem.
And the CuO$^-$ spectral thresholds are broadened by vibration and, in one case,
complicated by a long-lived triplet excited state; the comparison is not the
clean orbital-by-orbital match available here.

\section{Conclusions}\label{sec:concl}

We report inverse generalized Kohn--Sham calculations on polyatomic molecules
at a series of exact-exchange fractions $\alpha$ and a fixed near-exact CCSD
density, and compare the occupied spectra with measured vertical ionization
energies. To our knowledge this is the first such study. At $\alpha = 0$, 
where the GKS operator reduces to the Kohn--Sham operator, our results (like 
previous ones) are consistent with the expected equality between
the lowest VIE and negative of the highest occupied orbital energy
eigenvalue (within numerical errors 
of the order 0.1 eV).

 The central finding is that the Chong--Gritsenko--Baerends square-root
relation, $\varepsilon_k^{\alpha} + I_k \propto
\sqrt{\eHOMO^{\alpha} - \varepsilon_k^{\alpha}}$, observed for the Kohn--Sham
system, holds across the whole hybrid family. Its coefficient is linear in
$\alpha$, $c(\alpha) = c(0) - s\alpha$, with $c(0) = 1.185 \pm 0.017~\eVh$ and
$R^{2} \ge 0.9996$, and vanishes at $\alpha^{*} = 0.597 \pm 0.008$ (0.06 lower with the $\alpha=0$
eigenvalues of CGB), where the HOMO-referenced occupied levels satisfy $\varepsilon_k \approx -I_k$. The zero crossing is fixed by the core orbitals. In the valence, adding exact
exchange at fixed density does not improve the occupied eigenvalues, but leaves
the HOMO-referenced deviations unchanged or degrades them. The value of $\alpha^{*}$ is close to the $0.62$ exact-exchange fraction of the QTP17 global hybrid, whose
parameterization weights core ionization energies most heavily.

The slope $s = 1.987 \pm 0.041~\eVh$ agrees with the homogeneous-electron-gas
step-height coefficient $\Kg = 1.9932~\eVh$ and with the response-potential
coefficient $K_{\mathrm{resp}} = 1.972 \pm 0.040~\eVh$ obtained from
Ref.~\cite{GBB2003}. We conjecture that exact-exchange admixture removes the
response contribution to $c(\alpha)$ at this rate while the spectroscopic
contribution stays independent of $\alpha$, so that the whole of $dc/d\alpha$
is attributable to the response term.

All seven molecules are weakly correlated at their equilibrium geometries, and
whether these conclusions extend to the strongly correlated regime remains
open. The benchmark establishes the accuracy of the inversion where the answer
is known. The natural next target is the copper oxide anions, on which every
tested density functional fails, and to which the validated procedure can now be
applied.

\section*{Acknowledgments}

JPP acknowledges support from the National Science Foundation under Grant Nos.
DMR-2426275 and CHE-CTMC-2533416.

VS, BK and VG acknowledge the support of the Department of Energy, Office of
Science, through grant number DE-SC0022241, under the auspices of which the
inverse calculations were performed. This study used resources of the Oak Ridge
Leadership Computing Facility, a DOE Office of Science User Facility supported
under Contract DE-AC05-00OR22725, and of the NERSC Center, a DOE Office of
Science User Facility, under NERSC award BES-ERCAP0034270.

\noindent\textit{AI disclosure statement.} During the preparation of this manuscript,
the authors used Claude Opus 5.5 (Anthropic) to generate and debug the Python
code for the data analysis, tables and figures, to detect anomalies in the data
and text, to locate and summarize relevant publications, and to edit and draft
text. The authors reviewed, validated and critically assessed all AI-assisted
outputs and take full responsibility for the content of the manuscript.

\pagebreak
\begin{center}
\textbf{\Large Supplementary Information}
\end{center}

\setcounter{equation}{0}
\setcounter{figure}{0}
\setcounter{table}{0}
\setcounter{page}{1}
\setcounter{section}{0}
\makeatletter
\renewcommand{\theequation}{S\arabic{equation}}
\renewcommand{\thefigure}{S\arabic{figure}}
\renewcommand{\thetable}{S\arabic{table}}
\renewcommand{\thesection}{S\arabic{section}}
\renewcommand{\thepage}{S-\arabic{page}}

The bond lengths and angles of the seven molecules are given in Table~\ref{tab:geom}.

\begin{table}[htbp]\centering\small
\caption{Bond lengths and angles of the seven molecules studied.}\label{tab:geom}
\begin{tabular}{@{}lrr@{}}\toprule
Molecule & $r$ (\AA) & angle (deg) \\ \midrule
LiH & 1.5950 & --- \\
LiF & 1.5640 & --- \\
H$_2$O & 0.9578 & 104.48 \\
HF & 0.9170 & --- \\
CO & 1.1280 & --- \\
N$_2$ & 1.0980 & --- \\
F$_2$ & 1.4120 & --- \\
\bottomrule\end{tabular}\end{table}

Symmetry-required degenerate orbitals must be exactly degenerate, so any
splitting the inversion produces measures its numerical error directly.
Table~\ref{tab:split} gives these splittings. Entries reading $0.000$ lie below
the $0.001$~eV precision at which the eigenvalues are held and are not exact
degeneracies. Where levels cross, orbitals are assigned by symmetry rather than
by energy order: the $3\sigma_g$ level of F$_2$ rises above the $1\pi_u$ pair near
$\alpha=0.72$ and that of N$_2$ falls below it near $\alpha=0.95$, and assigning
by energy order would pair a $3\sigma_g$ eigenvalue with a $1\pi_u$ one and
report a spurious splitting of up to $0.92$~eV.

\begin{table}[htbp]\centering\small
\caption{Splitting of symmetry-required degenerate orbital pairs (eV). The exact
value is zero.}\label{tab:split}
\begin{tabular}{@{}llrrrrr@{}}\toprule
Molecule & pair & $\alpha=0$ & $0.25$ & $0.5$ & $0.75$ & $1.0$ \\ \midrule
LiF & $1\pi$ & 0.000 & 0.000 & 0.000 & 0.000 & 0.003 \\
CO & $1\pi$ & 0.001 & 0.000 & 0.000 & 0.000 & 0.000 \\
HF & $1\pi$ & 0.000 & 0.000 & 0.000 & 0.000 & 0.000 \\
N$_2$ & $1\pi_u$ & 0.000 & 0.000 & 0.000 & 0.000 & 0.000 \\
F$_2$ & $1\pi_g$ & 0.000 & 0.000 & 0.000 & 0.000 & 0.002 \\
F$_2$ & $1\pi_u$ & 0.001 & 0.000 & 0.000 & 0.000 & 0.002 \\
\bottomrule\end{tabular}\end{table}

Table~\ref{tab:resp} divides the response-potential matrix elements
$(\mathbf{M}^{-1}\bm{\varepsilon}^{\mathrm{resp}})_i$ tabulated in
Ref.~\cite{GBB2003_SI} by the abscissa
$\sqrt{\varepsilon_{\mathrm{HOMO}}-\varepsilon_i}$ of Eq.~(5) of the main text,
evaluated from the same source. The ratio is to be compared with the
electron-gas step-height coefficient
$K_{\mathrm{g}}=8\sqrt{2}/3\pi^{2}=1.9932~\mathrm{eV}^{1/2}$.

\begin{table}[htbp]\centering\small
\caption{Response-potential matrix elements of Ref.~\cite{GBB2003_SI} divided by
the abscissa of Eq.~(5) of the main text. All six entries are core orbitals.}\label{tab:resp}
\begin{tabular}{@{}llrrr@{}}\toprule
Molecule & MO & $\sqrt{\varepsilon_{\mathrm{HOMO}}-\varepsilon_i}$ & $(\mathbf{M}^{-1}\bm{\varepsilon}^{\mathrm{resp}})_i$ & ratio \\ \midrule
CO & $1\sigma$ & 22.49 & 45.16 & 2.008 \\
CO & $2\sigma$ & 16.29 & 30.97 & 1.901 \\
HF & $1\sigma$ & 25.54 & 50.65 & 1.983 \\
H$_2$O & $1a_1$ & 22.47 & 44.98 & 2.002 \\
HCN & $2\sigma$ & 16.23 & 31.67 & 1.951 \\
HCN & $1\sigma$ & 19.34 & 38.45 & 1.988 \\
\midrule
mean & & & & $1.972 \pm 0.040$ \\
\bottomrule\end{tabular}\end{table}

\begin{figure}[htbp]\centering
\includegraphics[width=0.72\textwidth]{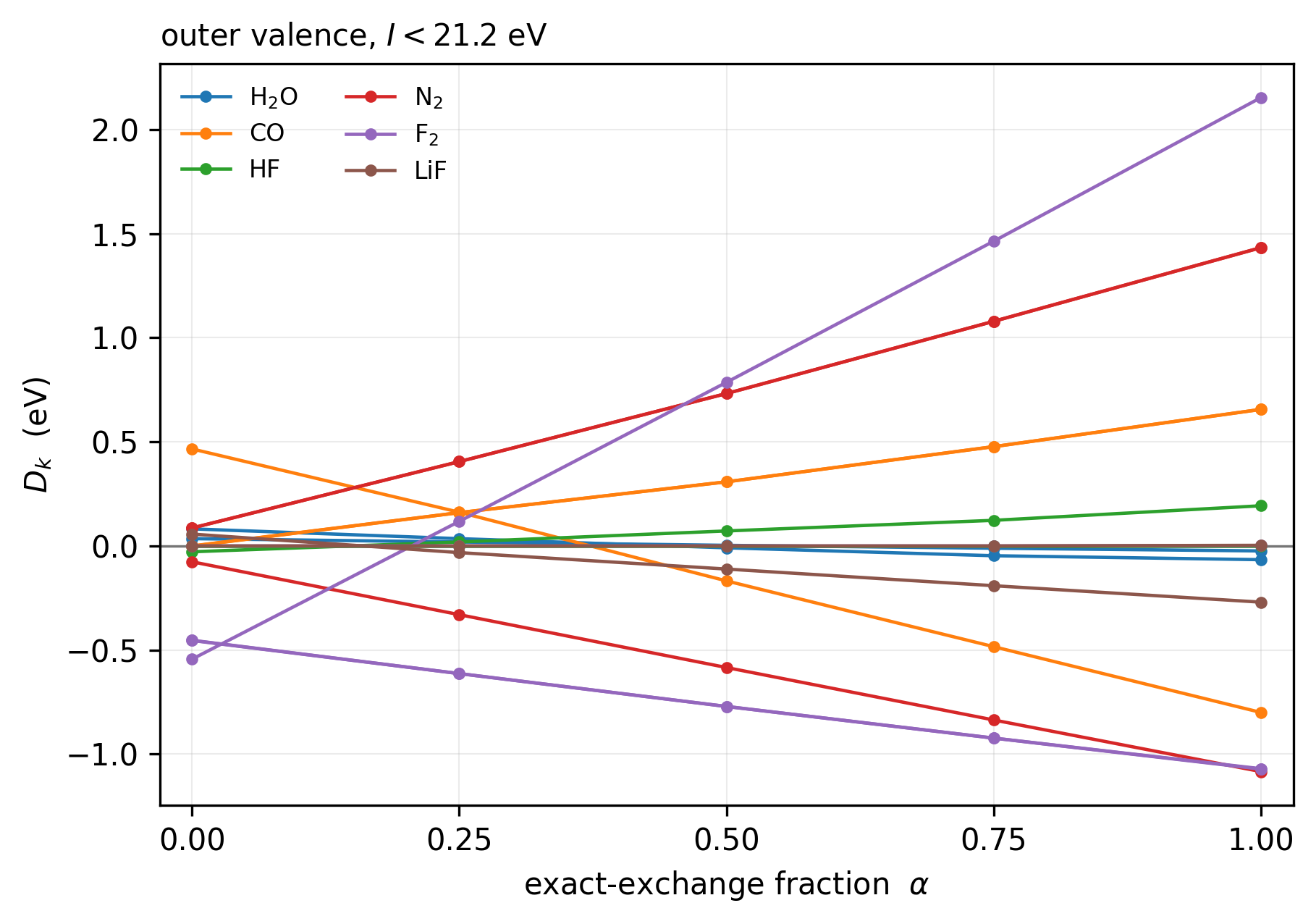}
\caption{HOMO-referenced deviation $D_k$ of the outer-valence orbitals
($I<21.2$~eV) against exact-exchange fraction $\alpha$, one line per orbital,
all molecules with outer-valence orbitals besides the HOMO, which excludes LiH. The HOMO itself is omitted, since $D_k=0$ there by
construction, as are the orbitals excluded from the fits. Compare Fig.~1 of the
main text, whose vertical scales are larger by factors of eight and sixty.}
\label{fig:ovDk}
\end{figure}

Figure~\ref{fig:ovDk} shows the HOMO-referenced deviation $D_k$ of the
outer-valence orbitals against the exact-exchange fraction, the companion to
Fig.~1 of the main text. Unlike the inner-valence and core deviations plotted
there, these start clustered near zero at $\alpha=0$ and fan outward as $\alpha$
grows, and almost none of them change sign. The aggregate behavior is a growth
of the spread rather than a displacement of the center: over the primary set the
mean deviation moves only from $+0.065$ to $+0.240$~eV while the standard
deviation grows from $0.14$ to $0.79$~eV. N$_2$ and F$_2$ carry most of that
growth, with spans of $1.24$ and $0.90$~eV over the full range of admixtures,
while H$_2$O and HF vary by $0.056$ and $0.123$~eV, within the $0.16$~eV error
floor.

\clearpage
\begin{thebibliography}{99}

\bibitem{KS1965} W.~Kohn and L.~J.~Sham,
\emph{Phys. Rev.} \textbf{140}, A1133 (1965).
\href{https://doi.org/10.1103/PhysRev.140.A1133}{doi:10.1103/PhysRev.140.A1133}

\bibitem{PPLB1982} J.~P.~Perdew, R.~G.~Parr, M.~Levy and J.~L.~Balduz,
\emph{Phys. Rev. Lett.} \textbf{49}, 1691 (1982).
\href{https://doi.org/10.1103/PhysRevLett.49.1691}{doi:10.1103/PhysRevLett.49.1691}

\bibitem{LPS1984} M.~Levy, J.~P.~Perdew and V.~Sahni,
\emph{Phys. Rev. A} \textbf{30}, 2745 (1984).
\href{https://doi.org/10.1103/PhysRevA.30.2745}{doi:10.1103/PhysRevA.30.2745}

\bibitem{AvB1985} C.-O.~Almbladh and U.~von Barth,
\emph{Phys. Rev. B} \textbf{31}, 3231 (1985).
\href{https://doi.org/10.1103/PhysRevB.31.3231}{doi:10.1103/PhysRevB.31.3231}

\bibitem{Baerends2022} E.~J.~Baerends, \emph{Phys. Chem. Chem. Phys.}
\textbf{24}, 12745 (2022).
\href{https://doi.org/10.1039/D2CP01585D}{doi:10.1039/D2CP01585D}

\bibitem{CGB2002} D.~P.~Chong, O.~V.~Gritsenko and E.~J.~Baerends,
\emph{J. Chem. Phys.} \textbf{116}, 1760 (2002).
\href{https://doi.org/10.1063/1.1430255}{doi:10.1063/1.1430255}

\bibitem{invGKS2026} V.~Subramanian, B.~Kanungo, V.~Khanna, P.~M.~Zimmerman and
V.~Gavini, arXiv:2609.15845 (2026).
\href{https://arxiv.org/abs/2609.15845}{arXiv:2609.15845}

\bibitem{Becke1993a} A.~D.~Becke, \emph{J. Chem. Phys.} \textbf{98}, 1372
(1993). \href{https://doi.org/10.1063/1.464304}{doi:10.1063/1.464304}

\bibitem{Becke1993b} A.~D.~Becke, \emph{J. Chem. Phys.} \textbf{98}, 5648
(1993). \href{https://doi.org/10.1063/1.464913}{doi:10.1063/1.464913}

\bibitem{KBP1996} W.~Kohn, A.~D.~Becke and R.~G.~Parr,
\emph{J. Phys. Chem.} \textbf{100}, 12974 (1996).
\href{https://doi.org/10.1021/jp960669l}{doi:10.1021/jp960669l}

\bibitem{Seidl1996} A.~Seidl, A.~G\"orling, P.~Vogl, J.~A.~Majewski and
M.~Levy, \emph{Phys. Rev. B} \textbf{53}, 3764 (1996).
\href{https://doi.org/10.1103/PhysRevB.53.3764}{doi:10.1103/PhysRevB.53.3764}

\bibitem{Garrick2020} R.~Garrick, A.~Natan, T.~Gould and L.~Kronik,
\emph{Phys. Rev. X} \textbf{10}, 021040 (2020).
\href{https://doi.org/10.1103/PhysRevX.10.021040}{doi:10.1103/PhysRevX.10.021040}

\bibitem{Gopidi2026} H.~R.~Gopidi, R.~Zhang, Y.~Wang, A.~Patra, J.~Sun, A.~Ruzsinszky,
J.~P.~Perdew and P.~Canepa, \emph{Phys. Rev. B} \textbf{113}, 165115 (2026).

\bibitem{Perdew2026} J.~P.~Perdew, T.~Lebeda, R.~Maniar, C.~Shahi, A.~Ruzsinszky and J.~Sun,
submitted to the Axel Becke special issue of \emph{J. Chem. Phys.},
edited by E.~Johnson and P.~Ayers.

\bibitem{GLLB1995} O.~V.~Gritsenko, R.~van Leeuwen, E.~van Lenthe and
E.~J.~Baerends, \emph{Phys. Rev. A} \textbf{51}, 1944 (1995).
\href{https://doi.org/10.1103/PhysRevA.51.1944}{doi:10.1103/PhysRevA.51.1944}

\bibitem{GBB2003} O.~V.~Gritsenko, B.~Bra\"ida and E.~J.~Baerends,
\emph{J. Chem. Phys.} \textbf{119}, 1937 (2003).
\href{https://doi.org/10.1063/1.1582839}{doi:10.1063/1.1582839}

\bibitem{Mezei2017} P.~D.~Mezei, G.~I.~Csonka and M.~K\'allay,
\emph{J. Chem. Theory Comput.} \textbf{13}, 4753 (2017).
\href{https://doi.org/10.1021/acs.jctc.7b00550}{doi:10.1021/acs.jctc.7b00550}

\bibitem{invDFT} V.~Subramanian, B.~Kanungo and V.~Gavini,
\emph{Comput. Phys. Commun.} \textbf{326}, 110218 (2026).
\href{https://doi.org/10.1016/j.cpc.2026.110218}{doi:10.1016/j.cpc.2026.110218}

\bibitem{DFTFE2022} S.~Das, P.~Motamarri, V.~Subramanian, D.~M.~Rogers and
V.~Gavini, \emph{Comput. Phys. Commun.} \textbf{280}, 108473 (2022).
\href{https://doi.org/10.1016/j.cpc.2022.108473}{doi:10.1016/j.cpc.2022.108473}

\bibitem{Kanungo2019} B.~Kanungo, P.~M.~Zimmerman and V.~Gavini,
\emph{Nat. Commun.} \textbf{10}, 4497 (2019).
\href{https://doi.org/10.1038/s41467-019-12467-0}{doi:10.1038/s41467-019-12467-0}

\bibitem{WuYang2003} Q.~Wu and W.~Yang,
\emph{J. Chem. Phys.} \textbf{118}, 2498 (2003).
\href{https://doi.org/10.1063/1.1535422}{doi:10.1063/1.1535422}

\bibitem{ACE2016} L.~Lin,
\emph{J. Chem. Theory Comput.} \textbf{12}, 2242 (2016).
\href{https://doi.org/10.1021/acs.jctc.6b00092}{doi:10.1021/acs.jctc.6b00092}

\bibitem{Baltzer1992} P.~Baltzer, M.~Larsson, L.~Karlsson, B.~Wannberg and
M.~Carlsson G\"othe, \emph{Phys. Rev. A} \textbf{46}, 5545 (1992).

\bibitem{Siegbahn1969} K.~Siegbahn \emph{et al.}, \emph{ESCA Applied to
Free Molecules} (North-Holland, Dordrecht, 1969).

\bibitem{Turner1971} D.~W.~Turner, C.~Baker, A.~D.~Baker and C.~R.~Brundle,
\emph{Molecular Photoelectron Spectroscopy} (Wiley-Interscience, New York,
1971).

\bibitem{Cavigliasso1999} G.~Cavigliasso and D.~P.~Chong,
\emph{J. Chem. Phys.} \textbf{111}, 9485 (1999).

\bibitem{MarieLoos2024} A.~Marie and P.-F.~Loos,
\emph{J. Chem. Theory Comput.} \textbf{20}, 4751 (2024).
\href{https://doi.org/10.1021/acs.jctc.4c00216}{doi:10.1021/acs.jctc.4c00216}

\bibitem{Ranasinghe2019} D.~S.~Ranasinghe, J.~T.~Margraf, A.~Perera and
R.~J.~Bartlett, \emph{J. Chem. Phys.} \textbf{150}, 074108 (2019).
\href{https://doi.org/10.1063/1.5084728}{doi:10.1063/1.5084728}

\bibitem{Jolly1984} W.~L.~Jolly, K.~D.~Bomben and C.~J.~Eyermann,
\emph{At. Data Nucl. Data Tables} \textbf{31}, 433 (1984).
\href{https://doi.org/10.1016/0092-640X(84)90011-1}{doi:10.1016/0092-640X(84)90011-1}

\bibitem{Carroll1974} T.~X.~Carroll, R.~W.~Shaw~Jr., T.~D.~Thomas, C.~Kindle and
N.~Bartlett, \emph{J. Am. Chem. Soc.} \textbf{96}, 1989 (1974).
\href{https://doi.org/10.1021/ja00814a001}{doi:10.1021/ja00814a001}

\bibitem{Berkowitz1979} J.~Berkowitz, C.~H.~Batson and G.~L.~Goodman,
\emph{J. Chem. Phys.} \textbf{71}, 2624 (1979).
\href{https://doi.org/10.1063/1.438619}{doi:10.1063/1.438619}

\bibitem{RoLinton1992} C.~Ro and R.~W.~Linton, \emph{Surf. Sci. Spectra}
\textbf{1}, 277 (1992).
\href{https://doi.org/10.1116/1.1247651}{doi:10.1116/1.1247651}

\bibitem{Eisenstadt1958} M.~Eisenstadt, G.~M.~Rothberg and P.~Kusch,
\emph{J. Chem. Phys.} \textbf{29}, 797 (1958).
\href{https://doi.org/10.1063/1.1744593}{doi:10.1063/1.1744593}

\bibitem{ZMP1994} Q.~Zhao, R.~C.~Morrison and R.~G.~Parr,
\emph{Phys. Rev. A} \textbf{50}, 2138 (1994).
\href{https://doi.org/10.1103/PhysRevA.50.2138}{doi:10.1103/PhysRevA.50.2138}

\bibitem{Shahi2024} C.~Shahi, R.~Maniar, J.~Ning, G.~I.~Csonka,
J.~P.~Perdew and A.~Ruzsinszky, \emph{J. Phys. Chem. A} \textbf{128}, 8628
(2024).
\href{https://doi.org/10.1021/acs.jpca.4c03640}{doi:10.1021/acs.jpca.4c03640}

\bibitem{Haiduke2018b} R.~L.~A.~Haiduke and R.~J.~Bartlett,
\emph{J. Chem. Phys.} \textbf{149}, 131101 (2018).
\href{https://doi.org/10.1063/1.5052442}{doi:10.1063/1.5052442}

\bibitem{Yang2020} X.~Yang, X.~Zheng and W.~Yang,
\emph{Front. Chem.} \textbf{8}, 588808 (2020).
\href{https://doi.org/10.3389/fchem.2020.588808}{doi:10.3389/fchem.2020.588808}

\bibitem{Haiduke2018} R.~L.~A.~Haiduke and R.~J.~Bartlett,
\emph{J. Chem. Phys.} \textbf{148}, 184106 (2018).
\href{https://doi.org/10.1063/1.5025723}{doi:10.1063/1.5025723}

\bibitem{Kanungo2024} B.~Kanungo, A.~D.~Kaplan, C.~Shahi, V.~Gavini and
J.~P.~Perdew, \emph{J. Phys. Chem. Lett.} \textbf{15}, 323 (2024).
\href{https://doi.org/10.1021/acs.jpclett.3c03088}{doi:10.1021/acs.jpclett.3c03088}

\end{thebibliography}

\begin{thebibliography}{99}
\bibitem{GBB2003_SI} O.~V.~Gritsenko, B.~Bra\"ida and E.~J.~Baerends,
\emph{J. Chem. Phys.} \textbf{119}, 1937 (2003).
\href{https://doi.org/10.1063/1.1582839}{doi:10.1063/1.1582839}
\end{thebibliography}
\end{document}